# Particle Simulation of Lower Hybrid Waves in Tokamak Plasmas

J. Bao[1, 2], Z. Lin[2], A. Kuley[2], Z. X. Wang[2] and Z. X. Lu[3, 4]

[1] Fusion Simulation Center and State Key Laboratory of Nuclear Physics and Technology, Peking University, Beijing 100871, China

[2] Department of Physics and Astronomy, University of California, Irvine, California 92697, USA

[3] Center for Momentum Transport and Flow Organization, UCSD, La Jolla, CA 92093, USA

## Abstract

Global particle simulations of the lower hybrid waves have been carried out using fully kinetic ions and drift kinetic electrons with a realistic electron-to-ion mass ratio. The lower hybrid wave frequency, mode structure, and electron Landau damping from the electrostatic simulations agree very well with the analytic theory. Linear simulation of the propagation of a lower hybrid wave-packet in the toroidal geometry shows that the wave propagates faster in the high field side than the low field side, in agreement with a ray tracing calculation. Electromagnetic benchmarks of lower hybrid wave dispersion relation are also carried out. Electromagnetic mode conversion are observed in toroidal geometry, slow waves are launched at the plasma boundary and converts to fast waves at the mode conversion layer, which is consistent with linear theory.

## 1. Introduction

Radio frequency (RF) waves are considered to be one of the most efficient tools for steady state operation of the tokamak, especially for heating and current drive [1-3]. We have developed a particle simulation model for the nonlinear RF physics by utilizing the existing physics capability, toroidal geometry and computational power of the gyrokinetic toroidal code (GTC) [4]. In this model, ions are treated as fully kinetic particles and electrons are treated as guiding centers by assuming that RF wavelength (i.e., LH wave) is much longer than electron gyro-radius [5, 6], which can handle the physics of the LH wave and ion Bernstein wave with a frequency smaller than the electron cyclotron frequency. We have carried out the particle simulations of the LH wave propagation in both the cylindrical and toroidal geometries for the first time [6]. As a benchmark exercise, we launch the LH wave from the tokamak edge to illustrate the difference of LH wave propagation between high and low field sides. We find that the poloidal mode number of the LH wave is not conserved due to the poloidal asymmetry in the toroidal geometry. The wave propagates faster in the high field side than the low field side, which is in agreement with well-known results obtained before by WKB simulations [7, 8]. The dispersion relations of the slow and fast branches

[4] Present address: Plasma Physics Laboratory, Princeton University, P.O. Box 451, Princeton, New Jersey 08543, USA

in LH frequency range are verified. The mode-conversion between slow and fast waves is observed in cylindrical and toroidal geometries, which is consistent with the theory [9].

## 2. Electrostatic simulation of LH wave

## 2.1. Electrostatic dispersion relation

An artificial antenna in GTC is used as a source to excite LH waves by adding the electric potential $\delta\phi_{ant}(r,\theta,\zeta,t) = \hat{\phi}(r)\cos(m\theta - n\zeta)\cos(\omega_{ant}t)$, where the subscript "ant" stands for "antenna". According to the driven resonant cavity theory, if the antenna frequency is equal to the eigen frequency of the system, the excited mode will have the maximal growth of the amplitude. In the simulation, we get the maximal growth of the LH wave amplitude as shown in figure 1(a) when the antenna frequency $\omega_{ant} = 0.98\omega_0$, where $\omega_0 = 3.39\omega_{LH}$ is the theoretical LH wave eigen-frequency. The difference is less than 2% between the GTC simulation and the theory. The radial structure of the LH wave from the GTC simulation is consistent with the analytical theory as shown in figure 1(b). Figures 1(c) and 1(d) are the corresponding poloidal plane of potential structures of the LH wave from theory and from GTC simulation, respectively. All the simulations in this paper are linear and the electron density fluctuation is effectively infinitesimal. It is seen that GTC simulation agrees with the analytical theory very well.

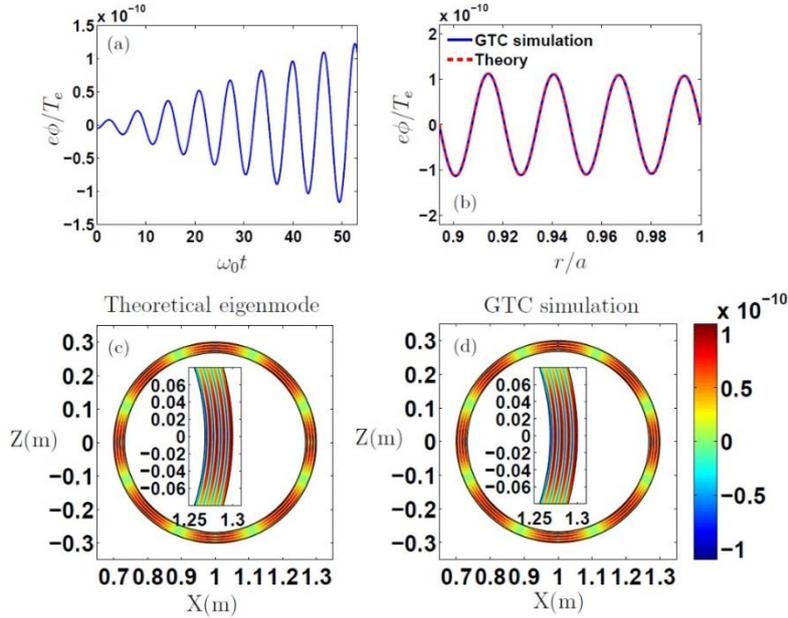

**Figure 1.** (a) Time evolution of the LH wave ($m = 4$, $n = 60$) amplitude excited by an artificial antenna in the cylinder, (b) radial profiles of LH waves from the GTC simulation and from the theory. (c) and (d) are poloidal mode structures of LH waves obtained from the theory and from the GTC simulation, respectively. The color scale represents the normalized electrostatic potential $\hat{\phi} = e\phi/T_e$.

For verifying the GTC simulation of the electron Landau damping, we focus on the verification of the linear electron Landau damping of a single mode in the cylindrical geometry. We choose the mode number randomly, this has no physical meaning with regard to the real experiments. Other values of mode numbers can also give the results in agreement with theory when $k_\parallel \ll k_\perp$ is satisfied. The simulation of the ($m=4$, $n=100$) LH wave with $k_\parallel a = 30.0$, $k_\perp a = 448.8$, $\omega_{pe} = 0.51\omega_{ce}$, $\omega_0 = 3.72\omega_{LH}$, $\xi_{i\perp} = \omega_0/k_\perp v_i = 6.67$ and $\xi_{e\parallel} = \omega_0/k_\parallel v_e = 2.33$ is performed in the cylinder ($a = 0.3m$ and $R = 1.0m$). The time evolution of the electrostatic potential $\phi$ of the LH wave in the simulation and a numerical fitting are shown in figure 3(a). The fitting function is defined as $\phi(t) = A\cos(\omega t + \alpha)e^{\gamma t}$, where $\omega$, $\gamma$ and $\alpha$ are the real frequency, damping rate, and initial phase of the LH wave, respectively. The damping rates obtained from the theory $\gamma_{ana} = -0.18\omega_{LH}$ and the simulation $\gamma_{simu} = -0.18\omega_{LH}$ are the same. Then we carry out the simulations with $\omega_{pe} = 0.51\omega_{ce}$ in different $\xi_{e\parallel}$ regimes, and compare simulation results of the damping rates with the theoretical calculation as shown in figure 3(b). Convergence tests indicate that 20 particles per cell are enough to simulate the linear electron Landau damping as shown in figures 3(c) and (d).

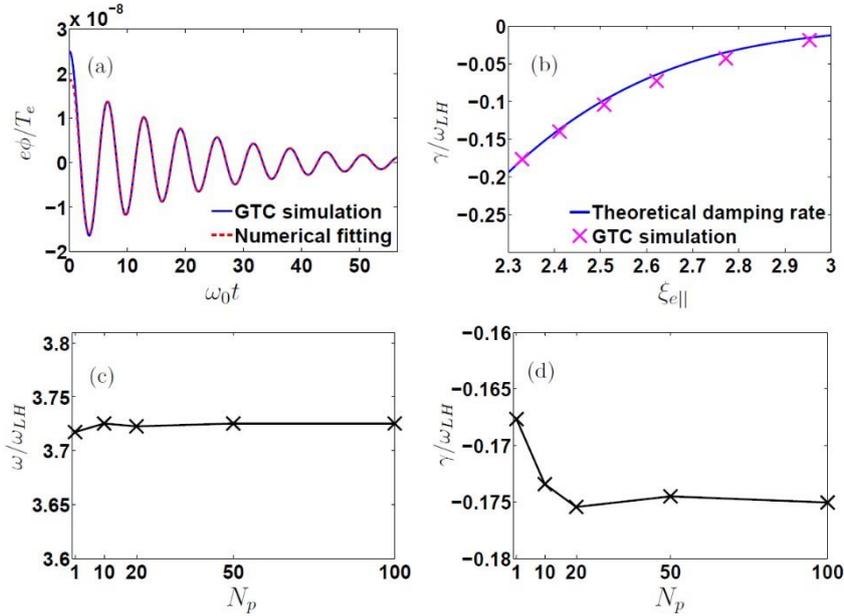

**Figure 3.** (a) Time history of the LH wave amplitude in the simulation. The dashed line is the numerical fitting. (b) Comparison of damping rates of LH waves obtained from the GTC simulation and the theoretical calculation in different $\xi_{e\parallel}$ regimes. (c) and (d) are the convergences of the number of particles per cell $N_p$ for LH wave frequency and damping rate, respectively.

## 2.2. LH wave propagation

In order to investigate the LH wave propagation in cylindrical plasmas, we apply an antenna as a source term to generate wave-packets at the outer boundary with all the information including the amplitude and initial phase. There is no energy loss of the LH wave during its propagation, since the plasma is cold. The wave-packets bounce back and forth between the inner and outer boundaries of the simulation domain to produce fluctuation patterns. Finally, the mode structure of the wave pattern is reconstructed in the simulation region. The radial profile of the wave pattern which is obtained from the GTC simulation agrees with the theoretical calculation in figure 4(b). The GTC simulation of the LH wave field is shown in figure 4(d), which agrees with the theoretical calculation shown in figure 4(c). Radial domain contains 10 wavelengths in the current simulation, which is smaller than real experiments.

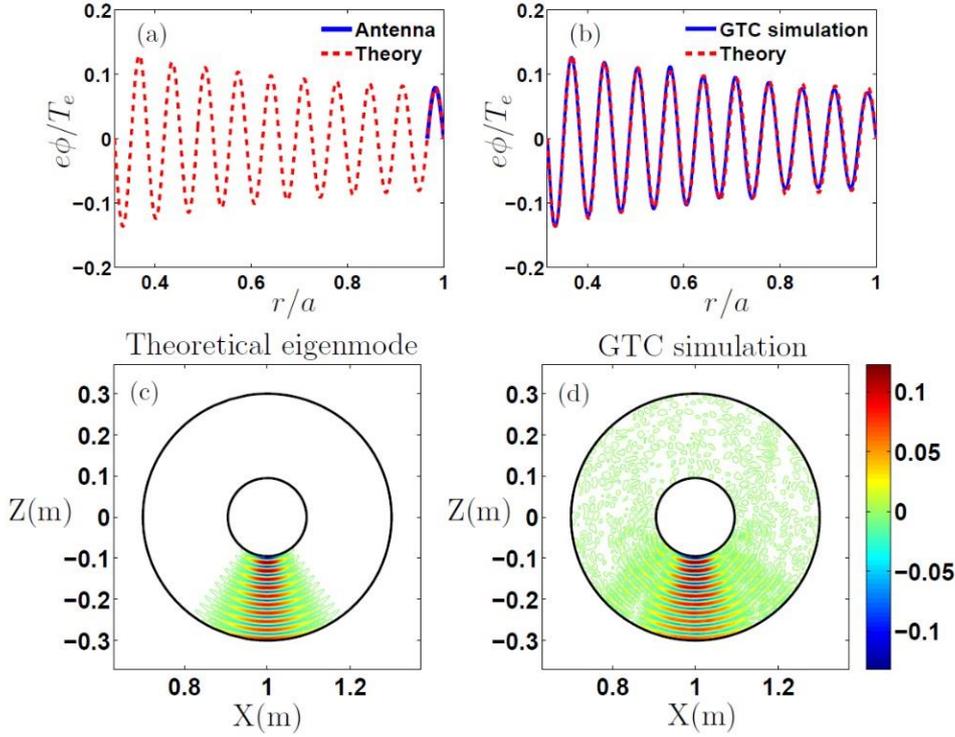

**Figure 4.** The red dash line in (a) is the theoretical wave pattern and the blue solid line in (a) is the radial profile of the artifical perturbation of the antenna. (b) is the comparison between radial profiles of wave patterns obtained from GTC simulation and the theory in the cylinder. (c) and (d) are electrostatic mode structures of LH wave patterns in the poloidal plane obtained from the theory and from the GTC simulation, respectively.

Finally, we carry out simulation of the LH wave propagation in the toroidal geometry. The structure of the wave-packet is formed by the coupling of different poloidal harmonics. In the toroidal geometry, the poloidal harmonic number $m$ of the wave-packet is not constant due to the poloidal asymmetry of **B** [6-8]. Thus the poloidal position of the wave-packet will change during its propagation as shown in figures 5(a) even though we launch the wave-packet perpendicular to the poloidal direction. A WKB simulation [8] gives the similar results of the LH wave propagation in the toroidal geometry as shown in figure 5(a). In the WKB simulation, two same rays are launched along the clockwise and counter-clockwise directions at the same time, respectively. The $m$ number decreases when the ray propagates from low to high field side as shown by the blue solid line in figure 5(b). The black dash line in figure 5(b) shows that the $m$ number increases when the ray propagates from high to low field side. Figure 5(c) and figure 5(d) show that the ray propagating in the high field side has a larger radial and poloidal group velocity than the low field side, respectively.

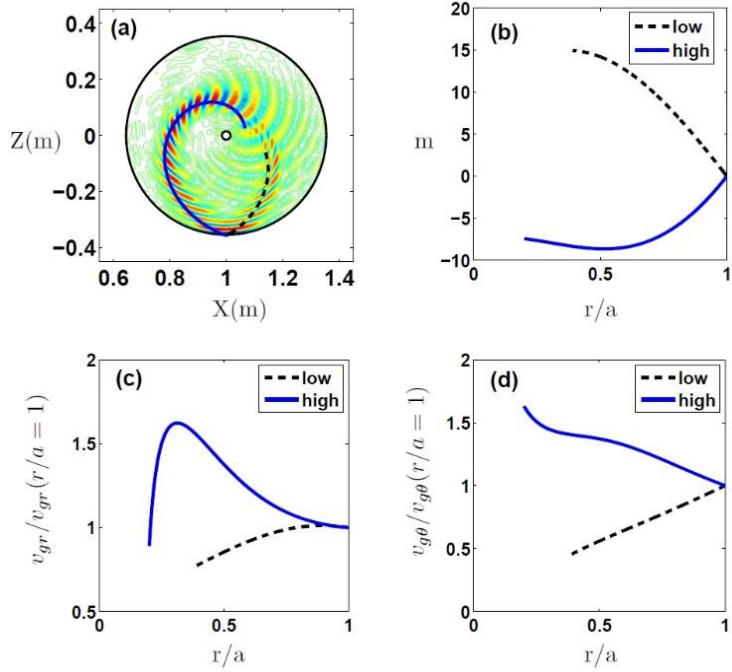

**Figure 5.** (a) WKB simulation of the ray trajectories is compared to GTC simulation in the toroidal geometry. (b)-(d) are the evolutions of the $m$ number, radial group velocity and the poloidal group velocity of the ray, respectively. The blue solid line represents the ray in the high field side, and the black dash line represents the ray in the low field side.

## 3. Electromagnetic simulation of LH wave

## 3.1 electromagnetic dispersion relation

Cold electromagnetic dispersion relation is $Sn_\perp^4 - Bn_\perp^2 + C = 0$, where $S$, $P$ and $D$ are the elements of Stix dielectric tensor, $B = (S - n_\parallel^2)(P + S) - D^2$ and $C = P\left[(S - n_\parallel^2)^2 - D^2\right]$. Parallel refractive index $n_\parallel = 2.088$ and frequency $\omega = 60.0\Omega_{ci}$, we choose slow and fast $n_\perp$ at $N_{e0} = 5 \times 10^{12} cm^{-3}$, and the turning point $N_{e0} = 8 \times 10^{12} cm^{-3}$. We use the antenna excitation and show the mode histories and structures [9].

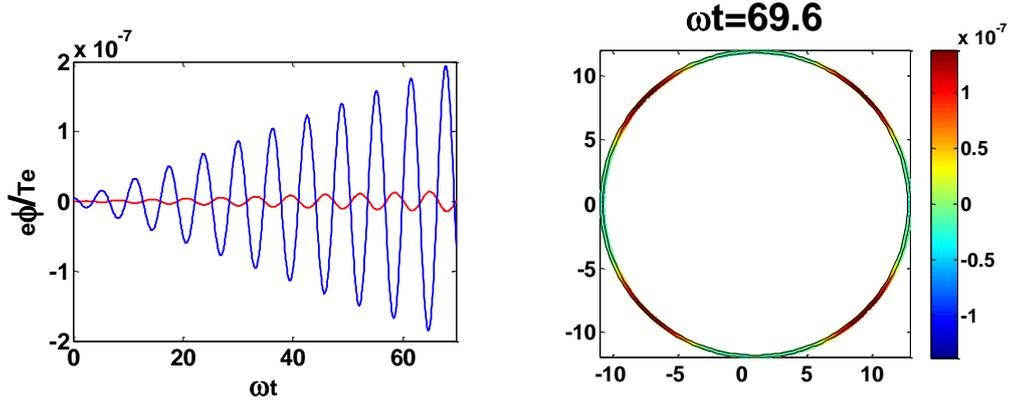

**Figure 6.** Show branch at $N_{e0} = 5 \times 10^{12}$, of which $k_\perp = 571.1 m^{-1}$. (a) is the mode history, and (b) is the mode structure.

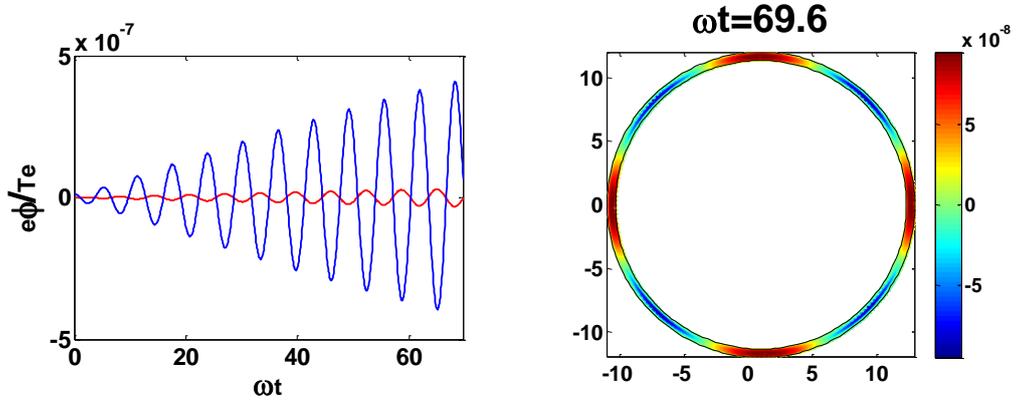

**Figure 7.** Fast branch at $N_{e0} = 5 \times 10^{12}$, of which $k_\perp = 193.2 m^{-1}$. (a) is the mode history, and (b) is the mode structure.

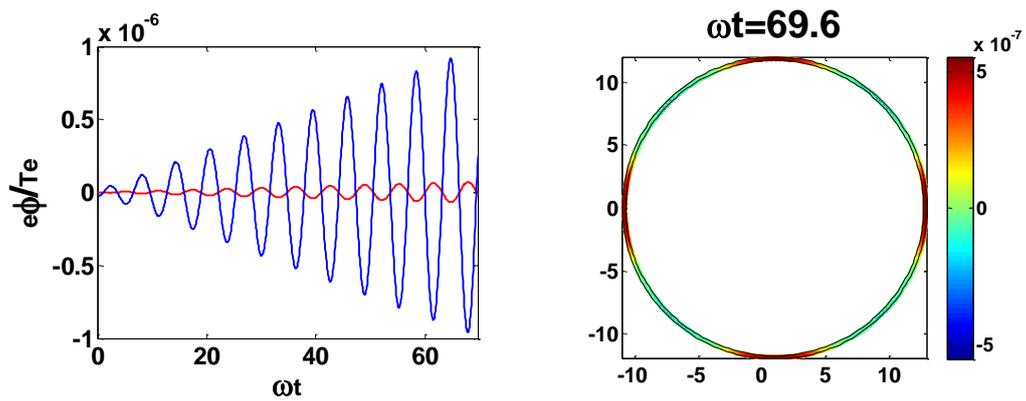

**Figure 8.** Turning point at $N_{e0} = 8 \times 10^{12}$, of which $k_\perp = 462.8 m^{-1}$. (a) is the mode history, and (b) is the mode structure.

## 3.2 Mode conversion of LH wave

Here, we launch the slow wave at $\theta = 0.25\pi$ at the plasma boundary, and the mode conversion happens when the accessibility condition is not satisfied as shown in figure 9. The radius-time 2-D plot of the slow wave part in figure 9 and its two-time-two-points correlation function are shown in figure 10, and the radius-time 2-D plot of the fast wave part in figure 9 and the corresponding correlation function are shown in figure 11. Figure 10(a) shows that the slow wave is launched at the plasma boundary, but its perpendicular phase velocity is outward to the plasma boundary as shown in figure 10(b). This is consistent with theory that the slow wave's perpendicular phase velocity and group velocity are opposite in sign. Whereas figure 11(b) shows that the mode converted fast wave's perpendicular phase velocity and group velocity are in the same direction, namely, propagating towards to the plasma boundary [9].

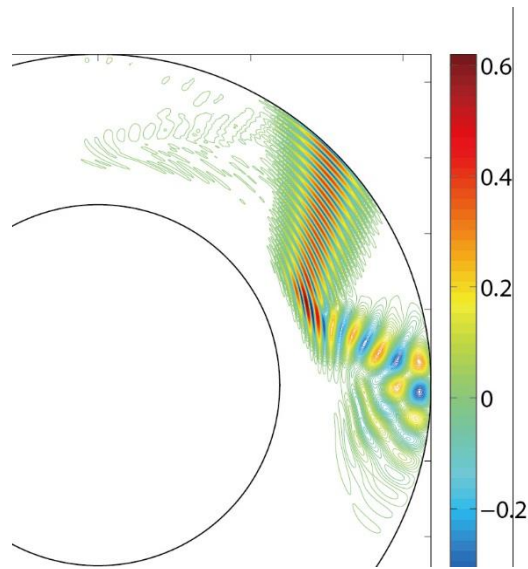

**Figure 9**. Mode conversion between slow wave and fast waves in toroidal geometry.

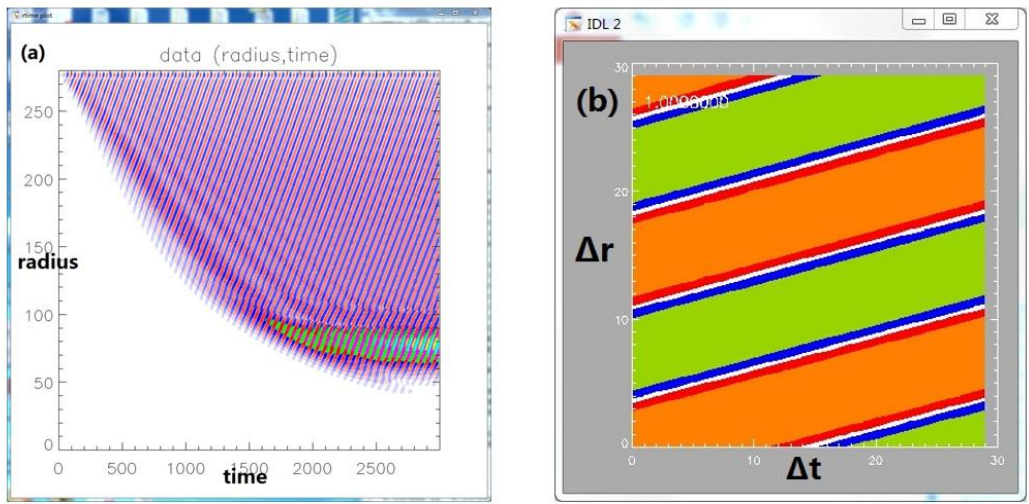

**Figure 10**. (a) 2-D radius-time plot of slow wave part in figure 9, and (b) two-time-two-points correlation function of slow wave.

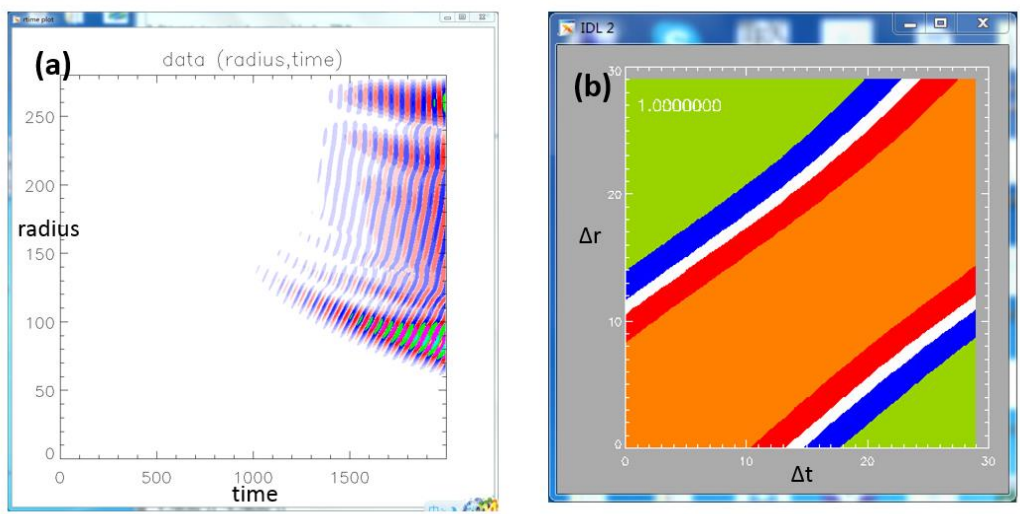

**Figure 11**. (a) 2-D radius-time plot of fast wave part in figure 9, and (b) two-time-two-points correlation function of fast wave.